\documentclass[prb,preprint]{revtex4}
\pdfoutput=1
\usepackage{graphicx}
\begin{document}


\title{Universality of density waves in \textit{p}-doped $\mathbf{La_2CuO_4}$  \linebreak and \textit{n}-doped $\mathbf{Nd_2CuO_{4+y}}$
\medskip }

\date{February 15, 2017} \bigskip

\author{Manfred Bucher \\}
\affiliation{\text{\textnormal{Physics Department, California State University,}} \textnormal{Fresno,}
\textnormal{Fresno, California 93740-8031} \\}

\begin{abstract}

The contribution of $O^{2-}$ ions to antiferromagnetism in $La_{2-x}Ae_xCuO_4$ ($Ae = Sr, Ba)$ is highly sensitive to doped holes. In contrast, the contribution of $Cu^{2+}$ ions to antiferromagnetism in $Nd_{2-x}Ce_xCuO_{4+y}$ is much less sensitive to doped electrons. The difference causes the precarious and, respectively, robust antiferromagnetic phase of these cuprates. The same sensitivities affect the doping dependence of the incommensurability of density waves, $\delta (x)$. In the hole-doped compounds this gives rise to a doping offset for magnetic and charge density waves, $\delta_{m,c}^p(x) \propto \sqrt{x-x_{0p}^N}$. Here $x_{0p}^N$ is the doping concentration where the N\'{e}el temperature vanishes, $T_N(x_{0p}^N) = 0$. No such doping offset occurs for density waves in the electron-doped compound. Instead, excess oxygen (necessary for stability in crystal growth) of concentration $y$ causes a different doping offset in the latter case, $\delta_{m,c}^n(x) \propto \sqrt{x- 2y}$. The square-root formulas result from the assumption of superlattice formation through partitioning of the $CuO_2$ plane by pairs of itinerant charge carriers. Agreement of observed incommensurability $\delta(x)$ with the formulas is very good for the hole-doped compounds and reasonable for the electron-doped compound. The deviation in the latter case may be caused by residual excess oxygen.

\bigskip
Keywords: High-temperature superconductors; Copper oxides; Stripes

\end{abstract}

\maketitle

\section{DENSITY WAVES IN HOLE-DOPED $\mathbf{La_{2}CuO_4}$}

Doping $La_{2}CuO_{4}$ with divalent alkaline-earth, $Ae$, substitutes ionized lanthanum atoms, $La \rightarrow La^{3+} + 3e^-$, by ionized
$Ae \rightarrow Ae^{2+} + 2e^-$ in the $LaO$ layers of the crystal. This causes electron deficiency (hole doping) of concentration $p = x$ in $La_{2-x}Ae_{x}CuO_4$ ($Ae = Sr, Ba$).
Each missing electron at the dopand site is replaced by an electron from an $O^{2-}$ ion, leaving an $O^-$ ion behind.
As will become clear, there is reason to assume that those $O^-$ ions reside in the $CuO_2$ planes sandwiched by the $LaO$ layers.
With respect to the parent compound, the $O^-$ ions electronically act as holes. 
Phenomenologically, it must be the doped holes that weaken, and rapidly destroy three-dimensional antiferromagnetism (3D-AFM) in $La_{2-x}Ae_{x}CuO_{4}$, in contrast to the robust 3D-AFM of electron-doped $Nd_{2-x}Ce_{x}CuO_{4}$.\cite{1}
The mechanism for the drastic difference of such dopand-dependent collapse of 3D-AFM in those compounds are still unknown. (One proposed mechanism for the contribution of $O^{2-}$ ions to 3D-AFM in high-transition-temperature copper oxides is in terms of loop currents.\cite{2} However, that approach remains controversial.\cite{3})

As experimentally observed at zero N\'{e}el temperature, $T_N(x_{0p}^N)=0$, the insulating 3D-AFM phase of $La_{2-x}Ae_{x}CuO_{4}$,  collapses at a small hole concentration $x_{0p}^N = x_{10} \equiv 2/10^2 = 0.02$ (see Fig. 1a) and a metallic, so-called ``pseudogap'' phase ensues. While the small amount of $O^{2-} \rightarrow O^{-}$ reduction destroys the 3D-AFM phase, the remaining $O^{2-}$ ions, together with the unaffected $Cu^{2+}$ spin magnetic moments, maintain 2D-AFM in the $CuO_{2}$ planes, sometimes called ``spin glass.''\cite{3}

Not only do the doped holes weaken and eventually destroy 3D-AFM, they also give rise to an incommensurate hole superlattice in $La_{2-x}Ae_{x}CuO_{4}$, as well as in $La_{1.6-x}Nd_{0.4}Sr_{x}CuO_{4}$ and $La_{1.8-x}Eu_{0.2}Sr_{x}CuO_{4}$. Thus the effect of hole doping on the crystals' magnetism is twofold: (i) a rapid collapse of their 3D-AFM and (ii) the emergence of a 2D incommensurate magnetic pattern. The latter can be regarded as an incommensurate magnetic dipole wave (MDW), also called spin density wave (SDW).\cite{4,5,6,7,8,9,10,11,12,13,14,15,16,17,18} From an electric point of view, the hole superlattice can be considered an incommensurate charge density wave (CDW).\cite{19,20,21,22,23,24,25,26,27,28,29,30,31,32,33,34,35} The MDWs and CDWs are thus related to one another by the constituting holes. Their unidirectional occurrence within domains is called ``stripes.''\cite{36}

With doping $x > x_{0p}^N$, the holes from the doping fraction  $x_{0p}^N$ are essentially localized to keep 3D-AFM suppressed. The remainder of the doping, $x - x_{0p}^N$, furnishes itinerant holes

\pagebreak 

\includegraphics[width=4.28 in]{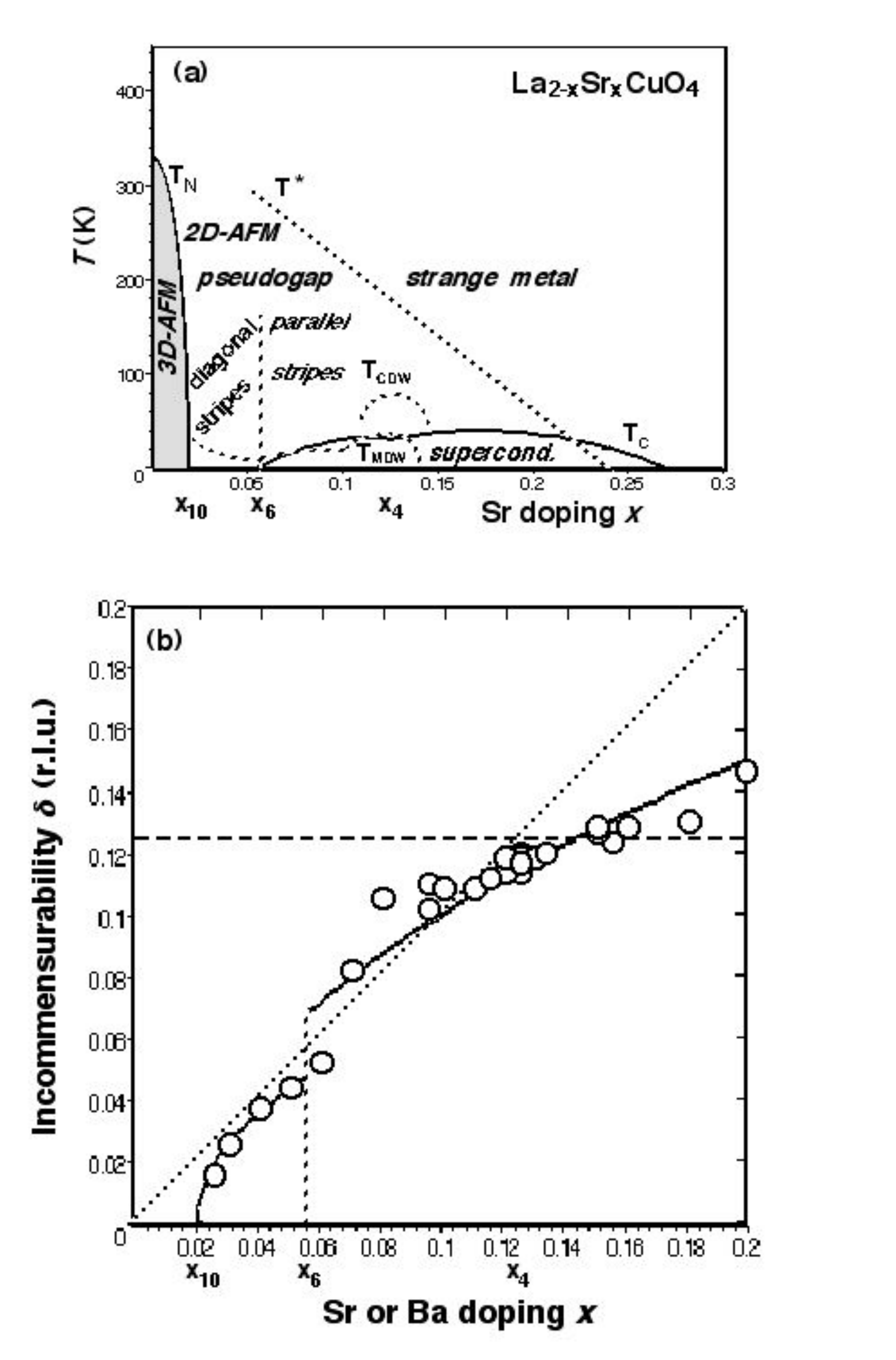}
\footnotesize 

\noindent FIG. 1. (a) Phase diagram of $La_{2-x}Sr_{x}CuO_{4}$ based on Refs. 7, 18, 24, 26, 38-43 with N\'{e}el temperature $T_N$, transition temperature of superconductivity $T_c$, pseudogap temperature $T^*$ (extrapolated to $T=0$) and temperatures $T_{CDW}$ and $T_{MDW}$ beneath which charge-density waves and, respectively, magnetic dipole waves are detected.
Special concentrations  $x_n = 2/n^2$, marked on the ordinate axis, correspond to commensurate doping of one $Sr$ atom per $na_{0}\times nb_{0}$ area in each $LaO$ plane.  
(b) Incommensurability of MDWs, $\delta_m^p = \delta$, and of CDWs $\delta_c^p = 2\delta$, in $La_{2-x}Ae_{x}CuO_{4}$ due to doping with $Ae = Sr$ or $Ba$, as well as in $La_{1.6-x}Nd_{0.4}Sr_{x}CuO_{4}$ and $La_{1.8-x}Eu_{0.2}Sr_{x}CuO_{4}$. Circles show data from neutron scattering or X-ray diffraction (Refs. 4-35). The broken solid curve is a graph of Eq. (1). Dotted slanted line $\delta = x$, dashed horizontal line at $\delta = 0.125 = 1/8$.
\pagebreak

\normalsize 
\noindent that segregate in the $CuO_{2}$ planes to an incommensurate superlattice of spacing $\lambda_c^p(x) \propto (x - x_{0p}^N)^{-1/2}$. It can be regarded as a static CDW of incommensurable wave number  $\delta_c^p(x) = 1/\lambda_c^p(x)$, first discovered in $La_{1.6-x}Nd_{0.4}Sr_{x}CuO_{4}$.\cite{4} The holes of the superlattice are located at the nodes of the incommensurable MDW of wavelength $\lambda_m^p(x)=2\lambda_c^p(x)$ based on defects in the nematic 2D-AFM pattern. In crystal domains large compared to the planar lattice constants, the combined CDWs-MDWs of incommensurability $\delta_c^p(x)=2\delta_m^p(x)$ are unidirectional (stripes). The dependence of the incommensurability on hole doping, in reciprocal lattice units (r.l.u.), is\cite{37}
\begin{equation}
\delta_{c,m}^p(x)  = w_{c,m} \frac{\Omega^{\pm}}{{4}}\sqrt {x - {x_{0p}^N}}  \; ,
\end{equation}

\noindent with a wave-kind factor $w_c=2$ or $w_m=1$ and a stripe-orientation factor $\Omega^{+}=\sqrt{2}$ for $x > x_6 \equiv 2/6^2  \simeq 0.056$ when stripes are parallel to the $a$ or $b$ axis, but $\Omega^{-} = 1$ for $x < x_6$ when stripes are diagonal. The density waves emerge at the doping level $x_{0p}^N=0.02$ where 3D-AFM collapses. The derivation of Eq. (1) is based on a partition of the $CuO_2$ plane by pairs of itinerant doped holes,\cite{37} incorporating the observed stripe orientation, here in tetragonal approximation of the lateral lattice constants, $a_0 = b_0$. The formula agrees to high accuracy with measured incommensurabilties of MDWs in hole-doped $La_{2-x}Ae_{x}CuO_{4}$, $La_{1.6-x}Nd_{0.4}Sr_{x}CuO_{4}$ and $La_{1.8-x}Eu_{0.2}Sr_{x}CuO_{4}$ by inelastic neutron scattering\cite{4,5,6,7,8,9,10,11,12,13,14,15,16,17,18} and of CDWs by hard X-ray diffraction or resonant soft X-ray scattering (RXS)\cite{19,20,21,22,23,24,25,26,27,28,29,30,31,32,33,34,35} (see Fig. 1b).

To put Eq. (1) in perspective, several comments are in order. 
(i) The experimentally observed relationship\cite{5} $\delta_c(x) = 2 \delta_m(x)$ is indicative of the mutual dependence of CDWs and MDWs, caused by doped holes. It implies that they \emph{always} exist together in copper oxides of 214 structure.
(ii) The observed transition from diagonal to parallel stripe orientation\cite{7} at doping $x_6$ coincides with the onset of superconductivity, $x_0^{SC} = x_6$. The reason for this agreement is still unknown.
(iii) The sensitivity of detection of MDWs by neutron scattering and of CDWs by hard X-ray diffraction or resonant soft X-ray scattering differs in several respects. In the parallel regime, $x > x_6$, the temperature profiles beneath which CDWs and MDWs are detected are dome-like with maxima at $x_4 = 1/8$, ranking $T_{CDW}(x) > T_c(x) > T_{MDW}(x) $ (see Fig. 1a). 
A possible reason for the dome-like shape of $T_{CDW}(x)$ and $T_{MDW}(x)$ could be, on the one hand, relatively large domain sizes for $x$ near the commensurate value $x_4$, antagonized by thermal agitation. 
On the other hand, the flanks of the domes could result from a diminution of domain size by structural instability due to proximity to the transition of diagonal/parallel stripes on the low-doping (left) side and proximity to the pseudogap/strange-metal transition on the higher-doping (right) side. In the diagonal regime, $x < x_6$, only MDWs have been observed, no CDWs (but possibly indications\cite{21} for diagonal CDWs in $La_{1.8-x}Eu_{0.2}Sr_{x}CuO_{4}$).
Concerning the different features of $T_{CDW}(x)$ and $T_{MDW}(x)$, it is conceivable that the diagonal pattern of 2D-AFM with wave vector $\mathbf{q}_{AFM} = (\frac{1}{2},\frac{1}{2})$ favors the domain size of diagonal MDWs. Conversely, the pattern of $Cu$-$O$ bonds parallel to the lateral crystal axes would favor the domain size of parallel CDWs.
(iv) The excellent agreement of data with Eq. (1) lends support to the underlying assumption that the doped holes reside in the $CuO_2$ plane. 
(v) Historically, the notion has been put forth\cite{6} that the doping dependence of the incommensurability of MDWs (and by implication of CDWs) in $La_{2-x}Ae_{x}CuO_{4}$ is essentially linear, $\delta_{c,m} \sim x$, up to 
$x \approx 0.12$, but levels off beyond to a value of $\delta_m(x) = \frac{1}{2} \delta_c(x) \approx 1/8 = 0.125$. The notion of such a ramp-like incommensurability profile, repeated many times in the literature, was helpful in the early days but is now rather outdated. As Fig. 1b shows, the data in the very low doping range indicate a parabolic (square-root like) dependence. 
Because of the \emph{break} of the square-root function at $x_6$ the branches of the solid curve skirt the dotted $\delta_m^/(x) = \frac{1}{2} \delta_c^/(x) = x$ line in the $0.04 < x < 0.10$ range from beneath and above, giving the false \emph{impression} of a linear dependence. The cluster of data near $x = 1/8$ clearly shows that $\delta_m(\frac{1}{8}) = \frac{1}{2} \delta_c(\frac{1}{8}) < 1/8$. Beyond $x \approx 0.12$ the data keep rising instead of leveling off. No \emph{explanation} for the putative ramp profile of $\delta_{c,m}(x)$ has ever been issued. In contrast, the broken square-root dependence of $\delta_{c,m}(x)$, Eq. (1), has been \emph{derived} by a partitioning of the $CuO_2$ planes by pairs of itinerant doped holes, combined with the observed change of diagonal/parallel stripe orientation at $x_6$. It is one of the very few simple relationships for the pseudogap phase.

\section{CHARGE DENSITY WAVES IN ELECTRON-DOPED $\mathbf{Nd_2CuO_4}$}

Pristine $Nd_{2}CuO_{4}$ has the same (214) crystal structure as $La_{2}CuO_{4}$. A minor difference occurs in $Ce$-doped $Nd_{2-x}Ce_{x}CuO_4$ for $x>x_6$ when oxygen ions in the $LnO$ layers that bracket the $CuO_2$ planes switch from apical positions above or beneath the $Cu^{2+}$ ions (T type) to positions apical to the $O^{2-}$ ions of the $CuO_2$ plane (T' type).\cite{1}
It is noteworthy that the T $\rightarrow$ T' transition at $x_6$ in the \textit{n}-doped 214 copper oxides coincides with the transition 

\pagebreak

\includegraphics[width=5.55in]{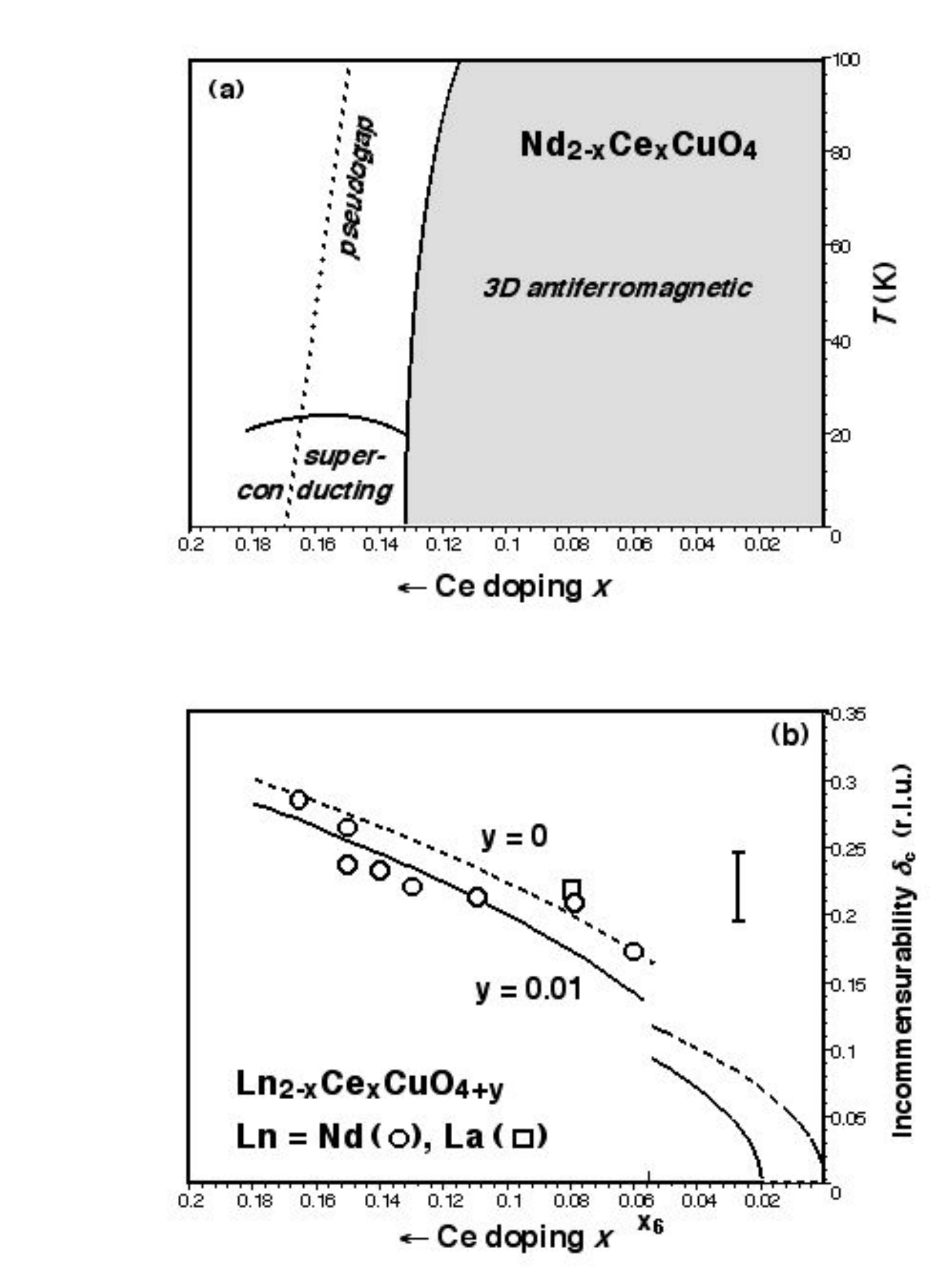}

\noindent FIG. 2. (a) Phase diagram of electron-doped $Nd_{2-x}Ce_{x}CuO_{4}$ after Ref. 44. (b) Observed incommensurability $\delta_c$ of CDWs in $Ln_{2-x}Ce_xCuO_{4+y}$ ($Ln = Nd$, crystal, circles; $Ln = La$, film, square)
from Refs. 45 and 46. The curves are graphs of Eq. (2) without excess oxygen ($y=0$, dashed line) and with excess oxygen of $y=0.01$ (solid line). An average error bar of the the data is shown to the right. The incommensurability of the corresponding MDW curve, $\delta_m(x) = \frac{1}{2} \delta_c(x)$, is not shown for sake of clarity.
\pagebreak

\noindent from diagonal to parallel stripes at $x_6$ in the \textit{p}-doped 214 compounds. Doping the parent crystal $Nd_{2}CuO_{4}$ with \emph{tetravalent} cerium substitutes ionized lanthanum atoms, $La \rightarrow La^{3+} + 3e^-$, by ionized $Ce \rightarrow Ce^{4+} + 4e^-$, causing electron surplus (electron doping) of concentration $n = x$ in $Nd_{2-x}Ce_{x}CuO_4$. The surplus electrons attach themselves to copper ions in the $CuO_2$ plane, $Cu^{2+} + \, e^- \rightarrow Cu^+$. Having a closed-shell electron configuration, the $Cu^+$ ions have no electron-spin magnetic moment, contrary to the unaffected $Cu^{2+}$ ions. This causes a weakening of AFM, although (for reasons unknown) at a much lesser rate than for hole doping.
Collapse of 3D-AFM occurs\cite{44} in electron-doped $Nd_{2-x}Ce_{x}CuO_4$ at  $x_{0n}^N \simeq 0.134$  (see Fig. 2a), in stark contrast to $x_{0p}^N = 0.02$ in hole-doped $La_{2-x}Ae_{x}CuO_4$, as mentioned.

The doped electrons in ${Nd_{2-x}Ce_{x}CuO_4}$ are free enough to segregate in the $CuO_2$ plane to an incommensurate superlattice, which can be regarded as a static CDW. Reducing a fraction, $n$, of $Cu^{2+}$ ions to spin-free $Cu^+$, gives rise to a weak MDW superimposed on the strong AFM background. The MDW is probably too weak to be detected with present technology as no MDWs have been observed in ${Nd_{2-x}Ce_{x}CuO_4}$.
Using for those density waves the same assumption as for doped holes---here, formation of a 2D superlattice that partitions the $CuO_2$ plane by pairs of doped \emph{electrons}---combined with the same stripe orientation, should then give the incommensurability $\delta_{c,m}^n(x)$ in square-root dependence on electron doping.

However, two circumstances need to be considered. (i) Since doped electrons don't frustrate the contribution of $O^{2-}$ ions to AFM, no doped electrons are exempted from CDWs to keep 3D-AFM suppressed---hence no doping offset $\Delta x_n$ in $\delta_{c,m}^n(x)$ analogous to $\Delta x_p = x_{0p}^N$ in Eq. (1).
(ii) It is well-known that electron-doped copper oxides need to be grown, for reasons of stability, in an oxygen atmosphere.\cite{1} As-grown $Ln_{2-x}Ce_{x}CuO_{4+y}$ ($Ln = Pr, Nd, Sm$) typically contain excess oxygen of $y \approx 0.03$ which has to be removed subsequently by sufficient annealing in an $Ar$ atmosphere\cite{47} This raises the possibility that the samples under consideration may still contain residual excess oxygen. Upon ionization, $O \rightarrow  O^{2-} - 2e^-$, excess oxygen ions reside in $Ln_{2-x}Ce_{x}CuO_{4+y}$ interstitially at apical positions above or beneath the $Cu^{2+}$ ions. The minus sign in the oxygen ionization balance accounts for the taking of two doped electrons (if present) by an interstitial oxygen atom (otherwise of two electrons from other atoms). 
This amounts, besides electron doping via $Ce$, to additional hole doping, $p = 2y$, which neutralizes a fraction of the doped-electron density, $\Delta n = -2y$, below the concentration of $Ce$, $n < x$. The consequence is a doping offset from excess oxygen affecting the incommensurability in electron-doped compounds, 
\begin{equation}
\delta_{c,m}^n(x) = w_{c,m} \frac{\Omega^{\pm}}{{4}}\sqrt{x - 2y}  \; ,  
\end{equation}

\noindent with $w_{c,m}$ and $\Omega^{\pm}$ as in Eq. (1).

Figure 2b shows incommensurabilities of CDWs in ${Nd_{2-x}Ce_{x}CuO_{4+y}}$ and ${La_{1.92}Ce_{0.08}CuO_4}$, observed with resonant X-ray scattering (RXS).\cite{44,45} The data fall close to the dashed curve of Eq. (2) for the case without excess oxygen ($y=0$) \emph{only} when $Ce$ doping is small or large, whereas the data in the middle range of doping fall beneath the dashed curve by about 15\%. However, for the case with $y=0.01$ oxygenation the data closely skirt the \emph{solid} curve. This raises the likelihood of residual excess oxygen in the samples. For a resolution of this possibility, enhanced analysis of excess oxygen would be valuable.

\section{SUMMARY AND REMARK}

Leaving aside the deviation in Fig. 2b for clarification by future experiment, the overall finding for 214 copper oxides is a universal \emph{square-root} dependence of density-wave incommensurability $\delta(x)$ on the doping of charge carriers---be they holes, be they electrons---quantitatively expressed by Eqs. (1) and (2). Their commonality is the underlying assumption that the $CuO_2$ plane is partitioned by pairs of itinerant charge carriers. 
The major difference between Eqs. (1) and (2) is the doping offset by $\Delta x_p = x_{0p}^N$ with $Ae$ (hole) doping but \emph{no} corresponding offset with $Ce$ (electron) doping. The difference is related on the one hand to the precarious sensitivity of the $O^{2-}$ ions' contribution to 3D-AFM when doped by holes, together with the necessity to keep 3D-AFM suppressed by a hole concentration $x_{0p}^N$. On the other hand it is related to the much lesser sensitivity of the $Cu^{2+}$ ions' contribution to 3D-AFM when doped by electrons. A minor difference between Eqs. (1) and (2) concerns the influence of excess oxygen---necessary for stability in crystal growth of ${Ln_{2-x}Ce_{x}CuO_{4+y}}$ ($Ln = Pr, Nd, Sm$)---by $\Delta x_n = -2y$.

As a refinement, the doping offset $\Delta x_p = x_{0p}^N$ in the radical of Eq. (1) may be treated as doping dependent, $\Delta x_p(x)$. When for $x \approx 1/8 = 0.125$ a value of $\Delta x_p(x) = 0.01$ (instead of $\Delta x_p(x) = x_{0p}^N = 0.02$) is used, agreement with experiment improves from very good (3\% deviation) to excellent (1\% deviation or less). In the framework of the above interpretation this could be understood as a lesser need to exempt holes from the CDW to keep 3D-AFM suppressed when overall more doped holes are present.

\bigskip \bigskip \bigskip

\centerline{ \textbf{ACKNOWLEDGMENTS}}

\noindent I thank Duane Siemens for discussions, suggestions, and critique. I also thank Preston Jones for help with LaTeX.


\begin{thebibliography}{47}


\bibitem{1} N. P. Armitage, P. Fournier, and R. L. Greene,  Rev. Mod. Phys. \textbf{82}, 2421 (2010).

\bibitem{2} C. M. Varma, Phys. Rev. B \textbf{55}, 14554 (1997); Phys. Rev. Lett. \textbf{83}, 3538 (1999).

\bibitem{3} J. M. Tranquada, G. Xu, and I. A. Zaliznyak, J. Mag. Mag. Mat. \textbf{350}, 148 (2014).

\bibitem{4} J. M. Tranquada, B. J. Sternlieb, J. D. Axe, Y. Nakamura, and S. Uchida, Nature \textbf{375}, 561 (1995).

\bibitem{5} J. M. Tranquada, J. D. Axe, N. Ichikawa, A. R. Moodenbaugh, Y. Nakamura, and S. Uchida
Phys. Rev. Lett. \textbf{78}, 338 (1997).

\bibitem{6} K. Yamada, C. H. Lee, K. Kurahashi, J. Wada, S. Wakimoto, S. Ueki, H. Kimura, Y. Endoh, S. Hosoya, G. Shirane, R. J. Birgeneau, M. Greven, M. A. Kastner, and Y. J. Kim, Phys. Rev. B \textbf{57}, 6165 (1998).

\bibitem{7} M. Fujita, K. Yamada, H. Hiraka, P. M. Gehring, S. H. Lee, S. Wakimoto, and G. Shirane, Phys. Rev. B \textbf{65}, 064505 (2002).

\bibitem{8} M. Matsuda, M. Fujita, K. Yamada, R. J. Birgeneau, Y. Endoh, and G. Shirane, Phys. Rev. B \textbf{65}, 134515 (2002).

\bibitem{9} O. J. Lipscombe, S. M. Hayden, B. Vignolle, D. F. McMorrow, and T. G. Perring, Phys. Rev. Lett. \textbf{99}, 067002 (2007).

\bibitem{10} M. Matsuda, M. Fujita, S. Wakimoto, J. A. Fernandez-Baca, J. M. Tranquada, and K. Yamada, Phys. Rev. Lett. \textbf{101}, 197001 (2008).

\bibitem{11} M. Fujita, M. Enoki, and K. Yamada, J. Phys. Chem. Solids \textbf{69}, 3167 (2008). 

\bibitem{12} A. T. R{\o}mer, P. J. Ray, H. Jacobsen, L. Udby, B. M. Andersen, M. Bertelsen, S. L. Holm, N. B. Christensen, R. Toft-Petersen, M. Skoulatos, M. Laver, A. Schneidewind, P. Link, M. Oda, M. Ido, N. Momono, and K. Lefmann
Phys. Rev. B \textbf{91}, 174507 (2015).

\bibitem{13} H. Jacobsen, I. A. Zaliznyak, A. T. Savici, B. L. Winn, S. Chang, M. H{\"u}cker, G. D. Gu, and J. M. Tranquada, Phys. Rev. B \textbf{92}, 174525 (2015).

\bibitem{14}M. Fujita, H. Goka, K. Yamada, J. M. Tranquada, and L. P. Regnault, Phys. Rev. B \textbf{70}, 104517 (2004).

\bibitem{15} S. R. Dunsiger, Y. Zhao, Z. Yamani, W. J. L. Buyers, H. A. Dabkowska, and B. D. Gaulin, Phys. Rev. B \textbf{77}, 224410 (2008).

\bibitem{16} M. H\"{u}cker, M. v. Zimmermann, G. D. Gu, Z. J. Xu, J. S. Wen, G. Xu, H. J. Kang, A. Zheludev, and J. M. Tranquada, Phys. Rev. B \textbf{83}, 104506 (2011).

\bibitem{17} Z. Xu, C. Stock, S. Chi, A. I. Kolesnikov, G. Xu, G. Gu, and J. M. Tranquada, Phys. Rev. Lett. \textbf{113}, 177002 (2014).

\bibitem{18} M. Kofu, S.-H. Lee, M. Fujita, H.-J. Kang, H. Eisaki, and K. Yamada, Phys. Rev. Lett. \textbf{102}, 047001 (2009).

\bibitem{19} M. v. Zimmermann, A. Vigliante, T. Niem\"{o}ller, N. Ichikawa, T. Frello, J. Madsen, P. Wochner, S. Uchida, N. H. Andersen, and J. M. Tranquada, Europhys. Lett. \textbf{41}, 629 (1998).

\bibitem{20} T. Niem\"{o}ller, N. Ichikawa, T. Frello, H. H\"{u}nnefeld, N.H. Andersen, S. Uchida, J.R. Schneider, and J.M. Tranquada, Eur. Phys. J. B \textbf{12}, 509 (1999).

\bibitem{21} M. H\"{u}cker, G. D. Gu, J. M. Tranquada, M. v. Zimmermann, H.-H. Klauss, N.J. Curro, M. Braden, and B. B\"{u}chner, Physica C \textbf{460-462}, 170 (2007).

\bibitem{22}J. Fink, V. Soltwisch, J. Geck, E. Schierle, E. Weschke, and B. B\"{u}chner, Phys. Rev. B \textbf{83}, 092503 (2011).

\bibitem{23} A. J. Achkar, F. He, R. Sutarto, J. Geck, H. Zhang, Y.-J. Kim, and D. G. Hawthorn, Phys. Rev. Lett. \textbf{110}, 017001 (2013).

\bibitem{24} T. P. Croft, C. Lester, M. S. Senn, A. Bombardi, and S. M. Hayden, Phys. Rev. B \textbf{89}, 224513, (2014).

\bibitem{25} V. Thampy, M. P. M. Dean, N. B. Christensen, L. Steinke, Z. Islam, M. Oda, M. Ido, N. Momono, S. B. Wilkins, and J. P. Hill, Phys. Rev. B \textbf{90}, 100510(R) (2014).

\bibitem{26} N. B. Christensen, J. Chang, J. Larsen, M. Fujita, M. Oda, M. Ido, N. Momono, E. M. Forgan, A. T. Holmes, J. Mesot, M. Huecker, and M. v. Zimmermann, ``Bulk charge stripe order competing with superconductivity
in $La_{2-x}Sr_xCuO_4 (x=0.12)$,'' arXiv:1404.3192

\bibitem{27} X. M. Chen, V. Thampy, C. Mazzoli, A. M. Barbour, H. Miao, G. D. Gu, Y. Cao, J. M. Tranquada, M. P. M. Dean, and S. B. Wilkins, Phys. Rev. Lett. \textbf{117}, 167001 (2016).

\bibitem{28} H. Kimura, H. Goka, M. Fujita, Y. Noda, K. Yamada, and N. Ikeda
Phys. Rev. B \textbf{67}, 140503(R) (2003).

\bibitem{29} P. Abbamonte, A. Rusydi, S. Smadici, G. D. Gu, G. A. Sawatzky, and D. L. Feng, Nature Phys. \textbf{1}, 155 (2005).

\bibitem{30} Y. J. Kim, G. D. Gu, T. Gog, and D. Casa
Phys. Rev. B \textbf{77}, 064520 (2008).

\bibitem{31} J. Kim, H. Zhang, G. D. Gu, Y. J. Kim, J. Supercond. Nov. Magn. \textbf{22}, 251 (2009).

\bibitem{32} M. H\"{u}cker, M. v. Zimmermann, Z. J. Xu, J. S. Wen, G. D. Gu, and J. M. Tranquada, Phys. Rev. B \textbf{87}, 014501 (2013).

\bibitem{33} M. P. M. Dean, G. Dellea, M. Minola, S. B. Wilkins, R. M. Konik, G. D. Gu, M. Le Tacon, N. B. Brookes, F. Yakhou-Harris, K. Kummer, J. P. Hill, L. Braicovich, and G. Ghiringhelli, Phys. Rev. B \textbf{88}, 020403(R) (2013).

\bibitem{34} V. Thampy, S. Blanco-Canosa, M. Garc{\'i}a-Fern{\'a}ndez, M. P. M. Dean, G. D. Gu, M. F\"{o}rst, T. Loew, B. Keimer, M. Le Tacon, S. B. Wilkins, and J. P. Hill, Phys. Rev. B \textbf{88}, 024505 (2013).

\bibitem{35} V. Khanna, R. Mankowsky, M. Petrich, H. Bromberger, S. A. Cavill, E. M\"{o}hr-Vorobeva, D. Nicoletti, Y. Laplace, G. D. Gu, J. P. Hill, M. F\"{o}rst, A. Cavalleri, and S. S. Dhesi, Phys. Rev. B \textbf{93}, 224522 (2016).

\bibitem{36} S. A. Kivelson, I. P. Bindloss, E. Fradkin, V. Oganesyan, J. M. Tranquada, A. Kapitulnik, and C. Howald, Rev. Mod. Phys. \textbf{75}, 1201 (2003).

\bibitem{37} M. Bucher, ``Superlattice origin of incommensurable density waves in $La_{2-x}Ae_xCuO_4 (Ae = Sr, Ba)$,'' arXiv:1308.1396v2

\bibitem{38} P. G. Radaelli, D. G. Hinks, A. W. Mitchell, B. A. Hunter, J. L. Wagner, B. Dabrowski, K. G. Vandervoort, H. K. Viswanathan, and J. D. Jorgensen, Phys. Rev. B \textbf{49}, 4163 (1994). 

\bibitem{39} B. Keimer, N. Belk, R. J. Birgeneau, A. Cassanho, C. Y. Chen, M. Greven, M. A. Kastner, A. Aharony, Y. Endoh, R. W. Erwin, and G. Shirane, Phys. Rev. B \textbf{46}, 14034 (1992).

\bibitem{40} H.-J. Julien, Physica B \textbf{329-333}, 693 (2003).

\bibitem{41}  T. Timusk and B. Statt, Rep. Prog. Phys. \textbf{62}, 61 (1999).

\bibitem{42} J. L. Tallon and J. W. Loram, Physica C \textbf{349}, 53 (2001).

\bibitem{43} M. R. Norman, D. Pines, and  C. Kallin, Adv. Phys. \textbf{54}, 715 (2005).

\bibitem{44} E. M. Motoyama, G. Yu, I. M. Vishik, O. P. Vajk, P. K. Mang, and M. Greven, Nature \textbf{445}, 186 (2007).

\bibitem{45} E. H. da Silva Neto, R. Comin, F. He, R. Sutarto, Y. Jiang, R. L. Greene, G. A. Sawatzky, and A. Damascelli, Science \textbf{347}, 282 (2015).

\bibitem{46} E. H. da Silva Neto, B. Yu, M. Minola, R. Sutarto, E. Schierle, F. Boschini, M. Zonno, M. Bluschke, J. Higgins, Y. Li, G.Yu, E.Weschke, F. He, M. Le Tacon, R. L. Greene, M. Greven, G. A. Sawatzky, B. Keimer, and A. Damascelli, Science Advances \textbf{2}:e1600782 (2016).

\bibitem{47} M. Lambacher, T. Helm, M. Kartsovnik, and A. Erb, Eur. Phys. J. \textbf{188}, 61 (2010). 

\end{thebibliography}
\end{document}